\documentclass[12pt]{article}
\usepackage{epsf}
\def\be{\begin{equation}}
\def\ee{\end{equation}}
\def\bea{\begin{eqnarray}}
\def\eea{\end{eqnarray}}

\begin{document}
\begin{titlepage}
\begin{center}
{\Large \bf William I. Fine Theoretical Physics Institute \\
University of Minnesota \\}
\end{center}
\vspace{0.2in}
\begin{flushright}
FTPI-MINN-12/30 \\
UMN-TH-3118/12 \\
September 2012 \\
\end{flushright}
\vspace{0.3in}
\begin{center}
{\Large \bf Pion transitions from $\chi_{cJ}$ to $\eta_c$
\\}
\vspace{0.2in}
{M.B. Voloshin  \\ }
William I. Fine Theoretical Physics Institute, University of
Minnesota,\\ Minneapolis, MN 55455, USA \\
School of Physics and Astronomy, University of Minnesota, Minneapolis, MN 55455, USA \\ and \\
Institute of Theoretical and Experimental Physics, Moscow, 117218, Russia
\\[0.2in]

\end{center}

\vspace{0.2in}

\begin{abstract}
The charmonium transitions from the $\chi_{cJ}$ states to $\eta_c$ with emission of one or two pions are considered. It is shown that the only processes of such type arising in the leading order in the QCD multipole expansion are the decays $\chi_{c1} \to \eta_c \pi^+ \pi^-$ and $\chi_{c0} \to \eta_c \pi^0$. The absolute rate of the latter decay, in spite of being isotopically suppressed, is by about an order of magnitude larger than that of the former, so that the branching fractions for the two processes are approximately equal, and are expected to be significantly below the recent upper bound reported by BESIII. The rates of both decays are also related to that of another similar transition $h_c \to J/\psi \pi^0$.
\end{abstract}
\end{titlepage}

Recently the BESIII experiment has established~\cite{bes} interesting upper bounds on the rates of two-pion transitions from the $\chi_{cJ}$ charmonium states to the spin-singlet ground state $\eta_c$, $\chi_{cJ} \to \eta_c \pi^+ \pi^-$. The  hadronic transitions between states of heavy quarkonium with emission of a pion pair as well as of isospin-violating transitions with emission of a single neutral pion attract a considerable and long-lasting interest as being related to understanding the QCD dynamics of heavy and light hadrons. The purpose of the present paper is to revisit the theoretical description of such processes within the multipole expansion in QCD~\cite{gottfried,mv78}, specifically in application to the transitions between $P$-wave and $S$-wave states of a heavy quarkonium, which may be helpful in assessing the possibilities for further experimental studies of such transitions. The main result to be argued here is that out of four allowed transitions~\footnote{The decays $\chi_{c0} \to \eta_c \pi^+ \pi^-$ and $\chi_{c1} \to \eta_c \pi^0$ are forbidden by parity.}, $\chi_{c2} \to \eta_c \pi^+ \pi^-$, $\chi_{c2} \to \eta_c \pi^0$, $\chi_{c1} \to \eta_c \pi^+ \pi^-$, and $\chi_{c0} \to \eta_c \pi^0$, only the latter two are contributed by the leading $E1-M1$ term of the QCD multipole expansion, so that both transitions from the $J=2$ state $\chi_{c2}$ should be strongly suppressed. Moreover, the absolute rate of the isospin breaking process $\chi_{c0} \to \eta_c \pi^0$ should be larger than that for the decay  $\chi_{c1} \to \eta_c \pi^+ \pi^-$ by approximately an order of magnitude, which essentially compensates for similar difference of the total widths between $\chi_{c0}$ and $\chi_{c1}$ and makes the branching fractions for the two transitions approximately equal:
\be
{\cal B}(\chi_{c0} \to \eta_c \pi^0) \approx {\cal B}(\chi_{c1} \to \eta_c \pi^+ \pi^-)~.
\label{bb}
\ee
Furthermore, the absolute rate of the transition $\chi_{c0} \to \eta_c \pi^0$ is simply related to that for a similar decay of the spin-singlet $P$-wave charmonium $h_c$:
\be
\Gamma(\chi_{c0} \to \eta_c \pi^0) = 3 \, \Gamma(h_c \to J/\psi \pi^0)~.
\label{gg}
\ee
One more remaining transition of the same origin, the decay $h_c \to J/\psi \pi \pi$, is strongly suppressed kinematically, and is unlikely to be of an immediate significance for experimental studies.

Neither of the discussed $1P \to 1S$ transitions in charmonium has been observed so far. (With a possible exception of $h_c \to J/\psi \pi^0$, whose sighting reported by the E760 experiment~\cite{e760} has not been confirmed.)
A theoretical expectation for the absolute values of the discussed decay rates is uncertain due to poor knowledge of the relevant quarkonium transition amplitude for the $1P \to 1S$ transition. Using as a benchmark the value of the same amplitude for the $2S \to 1P$ transition from the known rate of the decay $\psi' \to h_c \pi^0$~\cite{pdg},  one can estimate $\Gamma(\chi_{c0} \to \eta_c \pi^0)) \sim$~ (few keV), although the reliability of such estimate is not presently clear.

The description within the QCD multipole expansion of the considered here pion transitions closely parallels that originally suggested~\cite{mv86} (also in the review \cite{mvc}) for  similar processes in bottomonium: $\Upsilon(3S) \to h_b(1P) \pi^+ \pi^-$ and $\Upsilon(3S) \to h_b(1P) \pi^0$, and an experimental evidence for the latter decay has been recently reported~\cite{babarh}.  All these processes involve a transition with the change of both the total spin of the heavy hark-antiquark pair as well as its orbital momentum, so that they proceed through the interference of the $E1$ and $M1$ terms in the multipole expansion. In this respect these decays are quite different from the heavy quark spin-conserving processes, e.g. $\psi' \to J/\psi \pi \pi$, or $\chi_{bJ}(2P) \to \chi_{bJ}(1P) \pi \pi$, and using either of the latter for normalization~\cite{lk} may lead to erroneous predictions. 

All essential details of the calculation can be found in the literature, e.g. in the review \cite{mvc}. I briefly outline the reasoning here in order to make the discussion somewhat more self contained.
The $E1$ and $M1$ terms in the multipole expansion are described by the following terms in the effective Hamiltonian 
\be
H_{E1}=-{1 \over 2} \xi^a \, {\vec r} \cdot {\vec E}^a ~,
~~~H_{M1}= - {1 \over 2
\, m_c}\, \xi^a \, ({\vec \Delta} \cdot {\vec B}^a)~,
\label{cme}
\ee
where $\xi^a=t_1^a-t_2^a$ is the difference of the color generators
acting on the quark and antiquark (e.g. $t_1^a = \lambda^a/2$ with
$\lambda^a$ being the Gell-Mann matrices),  ${\vec r}$ is the vector
for relative position of the quark and the antiquark, ${\vec \Delta}=({\vec \sigma}_c
-{\vec \sigma}_{\bar c})/2$ is the
difference of the spin operators for the the quark and antiquark. Finally, ${\vec E}^a$ and $\vec B^a$ are the chromoelectric and
chromomagnetic components of the gluon field strength tensor. The assumed here normalization convention is that the QCD coupling $g$ is absorbed into the definition of the gluon field strength. 

The general expression for the amplitudes of the transition between the states $X$ and $Y$ of charmonium with emission of one or two pions can thus be written in a form factorized into the product of the quarkonium transition amplitude and the amplitude for production of the pions by the gluonic operator
\bea
&&A(X \to Y \pi^0) = {1 \over 32 m_c} \, \langle \pi^0 | E^a_i B_k^a |0 \rangle \, \langle Y  | \xi^b (r_i \, {\cal G} \, \Delta_k +  \Delta_k \, {\cal G} \, r_i )  \xi^b | X \rangle~, \nonumber \\
&&A(X \to Y \pi^+ \pi^-) = {1 \over 32 m_c} \, \langle \pi^+ \pi^- | E^a_i B_k^a |0 \rangle \, \langle Y  | \xi^b (r_i \, {\cal G} \, \Delta_k +  \Delta_k \, {\cal G} \, r_i ) \xi^b | X \rangle~,
\label{ga}
\eea
where ${\cal G}$ is the Green's function for propagation of the heavy quark pair in a color octet state.

The amplitude for the single $\pi^0$ production by the gluonic operator in Eq.(\ref{ga}) is determined~\cite{aanom1} by the chiral anomaly in QCD once the isospin breaking by the mass difference $m_d-m_u$ between the down and up quarks is taken into account:
\be
\langle \pi^0 | E^a_i B_k^a |0 \rangle = \delta_{ik} \, {2 \sqrt{2} \, \pi^2 \over 3 } \, {m_d-m_u \over m_d + m_u} \, f_\pi \, m_\pi^2~,
\label{ebp}
\ee
where $f_\pi \approx 130\,$MeV is the pion decay constant. The amplitude for the dipion production by the gluonic operator is evaluated by noticing that the the product $E^a_i B_k^a$ is proportional to the off-diagonal $(0 m)$ component of the gluon field energy-momentum tensor $\theta^G_{\mu \nu}$. The matrix element $\langle \pi \pi |\theta^G_{\mu \nu}| 0 \rangle$ has been studied by Novikov and Shifman~\cite{ns} in terms of the fraction $\rho_G$ of the pion momentum carried by gluons, and the relevant amplitude in Eq.(\ref{ga}) is found as
\be
\langle \pi^+ \pi^- | E^a_i B_k^a |0 \rangle = 2 \pi \alpha_s \rho_G \, \epsilon_{ikm} \, (E_1 p_{2m}+ E_2 p_{1m})~,
\label{eb2p}
\ee
where $E_1, E_2$ and $\vec p_1, \vec p_2$ are the energies and the momenta of the two pions. The estimated~\cite{ns} numerical value of the coefficient in this expression, $2 \pi \alpha_s \rho_G \approx  2 \div 2.5$ is in agreement with the measured behavior of sub-leading details (in particular the $D$-wave) in the transition $\psi' \to J/\psi \pi \pi$ (a discussion and further references can be found in Ref.~\cite{mvc}). 

The spatial structure of the matrix elements (\ref{ebp}) and (\ref{eb2p}) uniquely determines the selection rules for the heavy quarkonium amplitudes in the expressions (\ref{ga}). Namely, the $\delta_{ik}$ dependence of the matrix element in Eq.(\ref{ebp}) selects only the one-pion transitions between quarkonium states with the same $J$: $\Delta J=0$, while the antisymmetric in $i$ and $k$ dependence $\epsilon_{ikm}$ in Eq.(\ref{eb2p}) selects only the two-pion transitions with $|\Delta J| \le 1$. For this reason neither the decay $\chi_{c2} \to \eta_c \pi^0$, nor $\chi_{c2} \to \eta_c \pi \pi$ receives any contribution from the leading $E1-M1$ term in the multipole expansion, so that both these decays should be significantly suppressed. 

The amplitudes of all the allowed by the selection rules $1P \to 1S$ transitions can be written in terms of one radial overlap amplitude
\be
I_{1S,1P}={1 \over 18 \, m_c} \langle R_{1S} | r \, {\cal G}_P + {\cal G}_S \, r | R_{1P} \rangle~,
\label{ai}
\ee
where $R_{1S}$, $R_{1P}$ and ${\cal G}_{S,P}$ are the radial wave functions for quarkonium and the radial partial wave color-octet Green's function for the corresponding orbital momentum. The expressions for the transition amplitudes read as
\bea
&&A[h_c(\vec a_h) \to J/\psi(\vec a_\psi) \, \pi^0]=  {2 \, \sqrt{2} \, \pi^2 \over 3 } \, {m_d-m_u \over m_d + m_u} \, f_\pi \, m_\pi^2 \, I_{1S,1P} \, (\vec a_h \cdot \vec a_\psi)~,\nonumber \\
&&A(\chi_{c0} \to \eta_c \, \pi^0) ={2 \, \sqrt{2} \, \pi^2 \over  \sqrt{3}} \, {m_d-m_u \over m_d + m_u} \, f_\pi \, m_\pi^2 \, I_{1S,1P}~, \nonumber \\
&&A[\chi_{c1} (\vec a_1) \to \eta_c \, \pi^+ \pi^-] = 2 \, \sqrt{2} \, \lambda \, I_{1S,1P} \, \left [ E_1 \, (\vec p_2 \cdot \vec a_1) + E_2 \, (\vec p_1 \cdot \vec a_1) \right ]~,
\label{3amp}
\eea
where $\vec a_h$, $\vec a_\psi$ and $\vec a_1$ are the polarization amplitudes of the spin one particles $h_b$, $J/\psi$ and $\chi_{c1}$ respectively, and the notation $\lambda = \pi \alpha_s \rho_G $ is introduced. Numerically $\lambda \approx 1$.

Clearly, the unknown charmonium overlap integral $I_{1S,1P}$ cancels in the ratio of the rates of the decays listed in Eq.(\ref{3amp}), and one readily finds the relation (\ref{gg}) between the first two decay rates (the phase space in these two decays is essentially the same), and
\be
{\Gamma(\chi_{c0} \to \eta_c \, \pi^0) \over \Gamma (\chi_{c1}  \to \eta_c \, \pi^+ \pi^-)} = {4 \pi^2 \over \lambda \, \Phi} \, \left ( \pi^2 \, {m_d-m_u \over m_d + m_u} \, f_\pi \, m_\pi^2 \right )^2 \, p_{\pi^0} \approx {13.7 \over \lambda^2}~,
\label{rg2}
\ee
where $p_{\pi^0} \approx 413\,$MeV is the pion momentum in the transition $\chi_{c0} \to \eta_c \, \pi^0$ and $\Phi$ is the phase space integral for the decay $\chi_{c1}  \to \eta_c \, \pi^+ \pi^-$:
\be
\Phi=\int_{m_\pi}^{\Delta-m_\pi} \left ( E_1^2 \, p_1^2 + E_2^2 \, p_2^2 \right ) \, p_1 \, p_2 \, {\rm d} E_1 \approx (249\,{\rm MeV})^7~,
\label{phi}
\ee
with $\Delta=M(\chi_{c1})- M(\eta_c)$. The numerical value $(m_d-m_u)/(m_d+m_u) \approx 0.3$~\cite{gl} is used in the estimate in Eq.(\ref{rg2}). Naturally, at $\lambda \approx 1$ this estimate gives the relation in Eq.(\ref{bb}).

It can be mentioned that the rate of one more process described by the same overlap integral $I_{1S,1P}$, the decay $h_c \to J/\psi \pi \pi$, is strongly kinematically suppressed due to the significantly smaller mass difference between $h_c$ and $J/\psi$. Indeed, within the same approach the amplitude of this decay is given by
\be
A[h_c(\vec a_h) \to J/\psi(\vec a_\psi) \, \pi^+ \pi^-]=  2  \, \lambda \, I_{1S,1P} \, \epsilon_{ikm} (E_1 \,  p_{2m}  + E_2  p_{1m}) \, a_{hi} \, a_{\psi k}~,
\label{hpp}
\ee
and the rate can be related to that of $\chi_{c1}  \to \eta_c \, \pi^+ \pi^-$ as
\be
{\Gamma(h_c \to J/\psi \, \pi^+ \pi^-) \over \Gamma(\chi_{c1}  \to \eta_c \, \pi^+ \pi^-)} = {\Phi_1 \over \Phi} \approx 0.1~,
\label{hcr}
\ee
where $\Phi_1$ is the phase space integral as in Eq.(\ref{phi}), but with $\Delta = M(h_c)-M(J/\psi)$. Numerically $\Phi_1 \approx (181\,{\rm MeV})^7$. The value of the ratio (\ref{hcr}) and the relations (\ref{rg2}) and (\ref{gg}) imply, in particular, that the rate of the transition $h_c \to J/\psi \, \pi^+ \pi^-$ is approximately 40 times smaller than that for $h_c \to J/\psi  \pi^0$ in full agreement with the analysis of Ref.~\cite{mv86} of the transitions involving an $^1P_1$ quarkonium.

The absolute scale for the rates of the discussed decays depends on the unknown overlap integral $I_{1S,1P}$. It might be helpful, for an approximate orientation, to compare these decays with the transition $\psi' \to h_c \pi^0$, whose rate $\Gamma \approx 0.26\,$keV~\cite{pdg} is determined by a similar overlap integral $I_{1P,2S}$. One thus readily finds the ratio of the rates in terms of the ratio of the corresponding $I$:
\be
\Gamma(\chi_{c0} \to \eta_c \, \pi^0) = 3 \, \Gamma(h_c \to J/\psi \, \pi^0) = \Gamma (\psi' \to h_c \pi^0) \, {p_{\pi^0} \over k_{\pi^0}} \left | {I_{1S,1P} \over I_{1P,2S} }\right |^2 \approx (3.7\, {\rm keV}) \, \left | {I_{1S,1P} \over I_{1P,2S}} \right |^2~,
\label{12r}
\ee
where $k_{\pi^0} \approx 85\,$MeV is the pion momentum in the decay $\psi' \to h_c \pi^0$. Thus, if the overlap integrals for the $2S \to 1P$ and $1P \to 1S$ transitions are of similar value, the rate $\Gamma(\chi_{c0} \to \eta_c \, \pi^0)$ should be in the ballpark of few keV, and the branching fractions in Eq.(\ref{bb}) should be of the order of (few)$\times 10^{-4}$. 

As uncertain as the present estimate of the branching fractions (\ref{bb}) is, it still indicates that the recent experimental upper bound~\cite{bes} ${\cal B}(\chi_{c1} \to \eta_c \pi^+ \pi^-) < 0.32\%$ is by about an order of magnitude higher than a reasonable theoretical expectation. Also, the present estimates imply that it may be more advantageous for experimental studies to search for the decays with single pion: $\chi_{c0} \to \eta_c \, \pi^0$ and $h_c \to J/\psi \, \pi^0$. In spite of being suppressed by the small breaking of isospin by the light quark masses, these decays are enhanced by the contribution of the chiral anomaly (Eq.(\ref{ebp})) as has been pointed out long ago~\cite{mv86} and as it is indicated by the E760 results~\cite{e760} and by the data~\cite{babarh} on the transitions from $\Upsilon(3S)$ to $h_b(1P)$ in bottomonium. 

This work is supported, in part, by the DOE grant DE-FG02-94ER40823.

\end{document}